\documentclass[useAMS,usenatbib]{mn2e}

\newcommand{\ome}{$\omega$ Centauri}

\title[Spatial distribution in \ome]{The multiple stellar population in
\ome: spatial distribution and structural properties\thanks{Based on
Wide Field Image data collected at the European Southern Observatory,
La Silla, Chile during the observing programmes 62.L-0354 and
63.L-0439.}} 

\author[E. Pancino et al.] 
{E.~Pancino$^1$\thanks{E-email: pancino@bo.astro.it}, A.~Seleznev$^2$, 
F.~R.~Ferraro$^3$, M.~Bellazzini$^1$, G.~Piotto$^4$\\ 
  $^1$Osservatorio Astronomico di Bologna, via Ranzani 1, I-40127 
     Bologna, Italy\\  
  $^2$Astronomical Observatory, Urals State University, Lenin's 
     ave. 51, Ekaterinburg, 620083 Russia.\\
  $^3$Dipartimento di Astronomia, Universit\`a di Bologna, via Ranzani 
     1, I-40127 Bologna, Italy\\   
  $^4$Dipartimento di Astronomia, Universit\`a di Padova, vicolo 
     dell'Osservatorio 5, I-35122 Padova, Italy}  

\begin{document}

\date{Accepted 0000 December 00. 
      Received 0000 December 00; 
      in original form 0000 December 00} 
 
\pagerange{\pageref{firstpage}--\pageref{lastpage}} 
\pubyear{2003} 

\maketitle

\label{firstpage}


\begin{abstract} We present a detailed analysis of the spatial
distribution of the stellar populations in the Galactic globular
cluster \ome. Taking advantage of the large photometric catalog
published by \citet{p00}, we confirm that metal-rich populations have a
spatial distribution which is significantly different from the
metal-poor dominant population. In particular: {\it (i)} the different
sub-populations have different centroids and {\it (ii)} the metal-poor
population is elongated along the E-W direction, while the metal-rich
populations are oriented along the orthogonal direction, i.e., N-S. The
evidence presented here can partially explain the weird spatial
metallicity segregation found by \citet{j98}, and further supports the
hyphothesis that different sub-populations in \ome\  might have had
different origins. \end{abstract}



\section{Introduction}
\label{intro}

The peculiar nature of the Galactic globular cluster \ome\ has been
known and studied for more than forty years. Besides being the most
massive and luminous in the Milky Way, it is presently the only
globular cluster that shows a spread in its heavy elements content.
Recent findings have shown that {\it (i)} at least three primary
enrichment peaks do exist in this cluster, including the recently
discovered metal-rich component \citep{lee,p00}; {\it (ii)} an age
spread of 3--5 Gyrs seems to be required to explain the turnoff region
morphology \citep{hughes,hk00}; {\it (iii)} the chemical enrichment of
the metal-poor and intermediate stars is mainly due to the retention of
SNe~II and intermediate mass AGB stars ejecta \citep{smith,norris};
{\it (iv)} the metal-rich stars appear to have a lower
$\alpha$-enhancement, most probably due to SNe~Ia pollution
\citep{p02}.

All these pieces of evidence suggest that \ome\ could be an
intermediate object between normal globular clusters, which are unable
to retain any of the supernovae ejecta, and the dwarf spehroidal
galaxies (dSph), which are the smallest stellar systems capable of
self-enrichment. \ome\ could also be the remaining nucleus of a dwarf
galaxy that was stripped during its interaction with the Milky Way, in
possible analogy with the complex and still debated case of M~54 and
the Sagittarius dSph. The possibility that \ome\ comes from
``outside'' the Milky Way seems also required to explain its present
orbit \citep{dinescu}.

However, other clues complicate the picture. For example, the 
unusually high ellipticity of $\omega$~Cen, that has been demonstrated 
to be sustained by rotation \citep{merritt94}, is compatible with the
flattened shapes resulting from the merger of two globular clusters
\citep*{makino}, and anyway the long relaxation time
\citep{george,merritt97} grants that \ome\ is not completely relaxed
dynamically. Moreover, Norris et al. (1997) showed that only stars
with  [Fe/H]$\le-1.2$ in $\omega$~Cen do rotate, while the more
metal-rich  components show no evident sign of rotation. Pancino et al.
(2000) showed that, while the metal-poor population exhibits the well
known E-W elongation, the two metal-rich populations show a more
pronounced ellipticity, but with an elongation in the N-S direction.
Finally, Ferraro et al. (2002) have shown how the most metal-rich
population shows a different bulk proper motion with respect to the
whole cluster. These pieces of evidence point toward a different
dynamical origin for the various sub-populations in $\omega$~Cen,
possibly resulting from a major merger event in the cluster's past
history.


\begin{figure} 
\vspace{22cm} 
\includegraphics{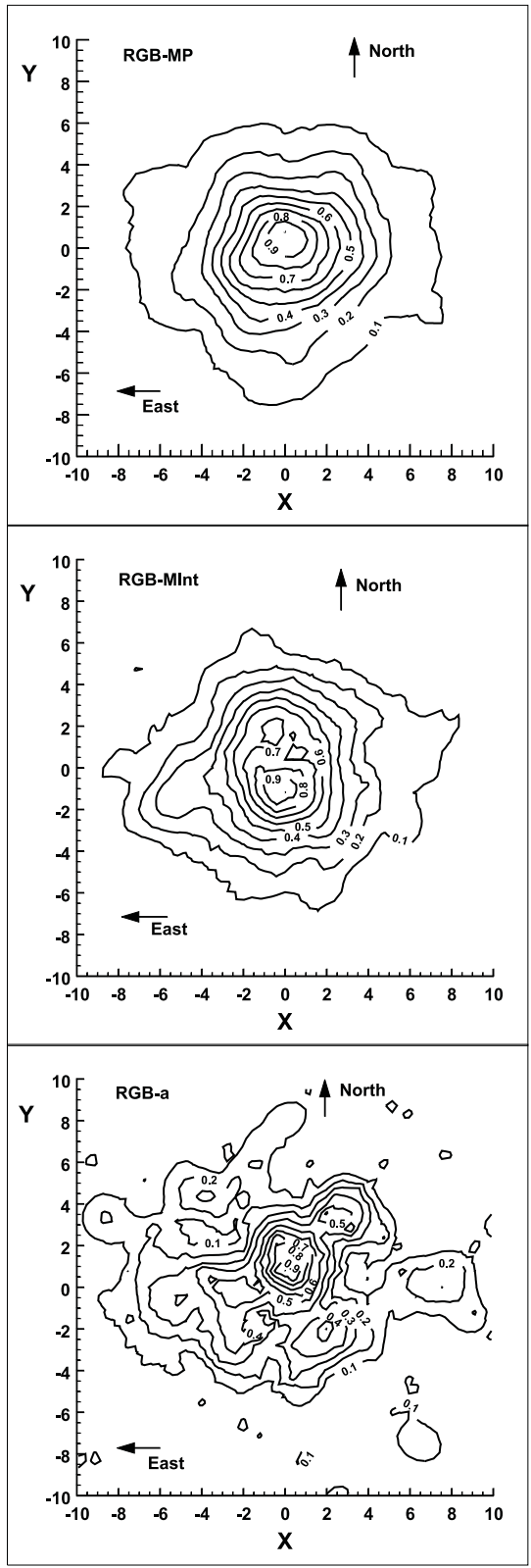}  

\caption{Isodensity contour lines for the three RGB samples defined in
the text: the RGB-MP (upper panel), the RGB-MInt (middle panel) and the
RGB-a (lower panel). The contour lines are normalized to their peak
density and are plotted in steps of 10\% of the peak density. In all
panels the axes show the distance from the cluster centre (700$\pm$20
pix, 1900$\pm$20 pix), in arcminutes. \label{dens}}

\end{figure} 


Given this framework, we have started a coordinated spectro-photometric
study of \ome\  \citep*[see][]{p00,p02,b01,f01,f02,o03}, specifically
devoted to the characterization of the sub-populations, and aimed at
understanding the origin and evolution of this complex stellar system.
In this paper, we exploit the large number of stars (more than 230,000
in total) and the wide area coverage ($33^{\prime}\times33^{\prime}$)
of the photometry published by \citet{p00}, to study in deeper detail
the structural properties of the red giants in \ome. 

The article is organized as follows. In Section~\ref{sample} we define
the photometric samples corresponding to the sub-populations of giants
in \ome; in Section~\ref{dens-par} we study the surface density
distributions and concentration of the three sub-samples; in
Section~\ref{elli-par} we measure the ellipticity, the centres
positions and the orientation of the three sub-populations; in
Section~\ref{segr-par} we comment on the metallicity segregation found
by \citet{j98}. Finally, in Section \ref{concl}, we summarize our main
results and discuss them in the framework of the present theories about
the formation and evolution of \ome.


\section{Samples definition} 
\label{sample} 

Using the metallicity information from the low resolution Ca triplet
survey by \citet{n96} and the morphology of the RGB from the
colour-magnitude diagram, Pancino et al. (2000, see their Fig.~2)
identified three sub-populations of RGB stars, with different average
metallicity and photometric properties. To analyse in detail the
structural properties of these sub-populations, we have extracted from
the \citet{p00} catalogue the following photometric samples (with
$B<16$~mag):

\begin{enumerate} 
 
\item{the RGB-MP sample, corresponding to the main, metal-poor peak
of the metallicity distribution, around [Ca/H]$\sim$--1.4 or 
[Fe/H]$\sim$--1.7. This population comprises $\sim 70\%$ of the
whole RGB population, and our photometric sample contains 2630
stars.} 

\item{the RGB-MInt sample, comprising the secondary, intermediate
metallicity peak around [Ca/H]$\sim$--1.0 or [Fe/H]$\sim$--1.2,
together with the long, extended tail to higher metallicities. This
sub-population accounts for $\sim 25\%$ of the whole RGB population,
and our sample contains 816 stars.}

\item{the RGB-a sample, the newly discovered metal-rich population
that comprises $\sim 5\%$ of the RGB stars and has a metallicity of
[Ca/H]$\sim$--0.5 and [Fe/H]$\sim$--0.6 \citep{p02}. Our photometric
sample contains 128 stars. Although this last sub-population has many
fewer stars, it is the most numerous sample presently at our disposal,
and its size could be significantly increased only when the RGB-a
counterparts in other evolutionary phases (i.e., horizontal branch,
sub-giant brach and main sequence) will be identified.}

\end{enumerate}


\section{Surface density distributions} 
\label{dens-par} 

We computed the surface density distributions of the three
sub-populations defined in Section~\ref{sample}, using a fixed kernel
estimator algorithm \citep{s86,merritt94,s98}, with a kernel half-width
of 500 pixels\footnote{In what follows, it is useful to bear in mind
that the WFI scale is 0.238 arcseconds per pixel. Thus, 100 pixels
correspond approximately to 24 arcseconds.} and a grid of 100 pixel
cells. The resulting surface density plots are shown in
Figure~\ref{dens}, where the isodensity contour lines shown are
normalized to the maximum density of each distribution, in steps of
10\%. From now on, we will refer to each isodensity level using the
fraction of the peak value, i.e., 0.3 for the isodensity level
corresponding to 30\% of the peak value. The peak density values are:
48.3~stars~arcmin$^{-2}$ for the RGB-MP, 14.6~stars~arcmin$^{-2}$ for
the RGB-MInt and 2.3~stars~arcmin$^{-2}$ for the RGB-a.

A first, qualitative comparison of the three distributions shown in
Figure~\ref{dens} reveals the following general facts:
 
\begin{enumerate} 
 
\item{The RGB-MP population is clearly elongated along the E-W
direction, reflecting the well known elliptical shape of the whole
system. The main peak position is consistent with the cluster centroid
estimated by \citet{p00}.}

\item{Both the RGB-a and the RGB-MInt populations have perturbed
isodensity contour lines, showing structures similar to bubbles, shells
and/or tails. While in the case of the RGB-a the complexity increases
in the outer parts, where it is almost certainly due to low number
statistics, in the case of the RGB-MInt we find complex structures in
the central parts, where data points are more numerous.}

\item{The RGB-MInt population is clearly elongated along the N-S
direction in the inner parts, while in the external parts it seems
to become elongated in the E-W direction. The main peak lies south
of the RGB-MP peak, with a possible secondary peak north of it,
that gives an evident asymmetric shape to the distribution.} 

\item{The RGB-a population is also elongated along the N-S
direction in its inner parts; its main peak lies north of the
RGB-MP peak.}

\end{enumerate}


\begin{table} 

\caption{Results of the two-dimensional generalization of the K-S 
test for the three sub-populations. The first column shows the
populations that are actually compared, the second shows the
maximum difference in the cumulative distributions $D$, while the
third column shows the derived probability $P$ that the two
populations are drawn from the same parent distribution.}

\label{tab_prob2}  
\begin{tabular}{lll}  
{\bf Population} & $D$ & $P$ \\  
\hline  
RGB-a vs. RGB-MP    & 0.171 & 0.012 \\  
RGB-MInt vs. RGB-MP & 0.068 & 0.039 \\  
RGB-MInt vs. RGB-a  & 0.117 & 0.225 \\  
\hline  
\end{tabular}  
\end{table} 


A simple monodimensional Kolmogorov-Smirnov test \citep[K-S, see
e.g.,][]{numerical} as a function of a radial coordinate is not
sufficient to properly assess the significance of these features. We
thus used a two-dimensional generalization of the K-S statistical
test, which was proposed and investigated with Monte-Carlo experiments
by \citet{ff87}, as a variant of an earlier idea by \citet{p83}. This
test, similarly to the usual K-S test, quantifies the probability $P$
that two (two-dimentional) distributions are extracted from the same
parent distribution, using a more sophisticated definition of the
maximum difference $D$ in the cumulative distributions.

The results are summarized in Table~\ref{tab_prob2}. The very low
probabilities obtained in the comparison of the RGB-MP population
with both the RGB-MInt and the RGB-a ensures us that their spatial
distributions are significantly different, i.e., they cannot be
drawn from the same parent distribution. On the other hand, the
probability obtained in the comparison between the RGB-MInt and the
RGB-a populations confirms that they are not significantly
different from each other.


\begin{table} 

\caption{For each sub-population (column~1), the equivalent radius, in
pixels, of the ellipses used to approximate the 90\% isodensity
contour lines (close to the peak) is shown in column~2 (r$_{e,90\%}$)
and its ratio with the radius of the RGB-MP population in column~3
(R$_{90\%}$). Columns~4 and 5 show the corresponding values
(r$_{e,60\%}$ and R$_{60\%}$) for the ellipses used to approximate the
60\% isodensity contour level (close to half maximum).}

\label{conc} 
\begin{tabular}{lllll} 
{\bf Population} & r$_{e,90\%}$ & R$_{90\%}$ & r$_{e,60\%}$ & R$_{60\%}$ \\ 
\hline 
RGB-MP   & 213.8$\pm$6.5  & 1.00 & 578.9$\pm$13.5 & 1.00 \\ 
RGB-MInt & 169.3$\pm$15.9 & 0.79 & 601.7$\pm$7.6  & 1.04 \\ 
RGB-a    & 66.2$\pm$5.5   & 0.31 & 323.1$\pm$5.0  & 0.56 \\ 
\hline 
\end{tabular} 
\end{table} 


We can also notice from Figure~\ref{dens} that the RGB-a population
appears {\em more} concentrated than the RGB-MP one, while we cannot
say much about the RGB-MInt population, which has a complicated shape
in its inner isodensity contour lines. A more quantitative evaluation
confirms this impression: Table~\ref{conc} reports the equivalent
radii\footnote{The equivalent radius for an ellipse of semi-axes $a$
and $b$ is defined as $r_e=\sqrt{ab}$, i.e., the radius of a circle
with the same area.} of the three sub-populations, calculated with the
ellipse parameters derived in Section~\ref{elli-par}, for the 0.9 and
0.6 isodensity contour lines. The RGB-a equivalent radii are $\sim$2-3
times smaller than the RGB-MP correspondent radii.


\section{Ellipticity} 
\label{elli-par} 
  
\begin{figure} 
\vspace{14.5cm} 
\includegraphics{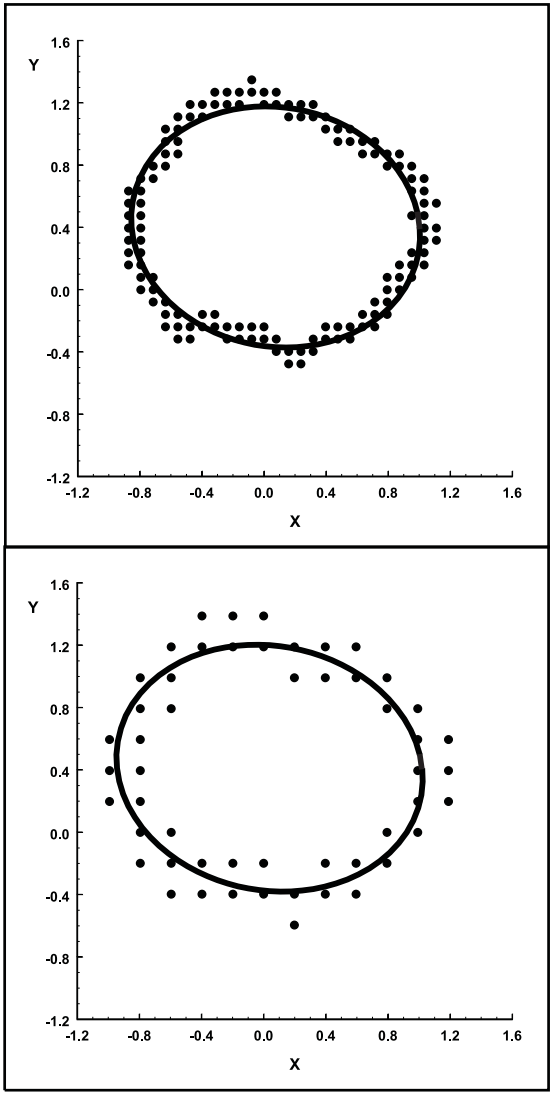} 

\caption{Example of the approximation of the 0.9 isodensity contour
line with an ellipse, for the RGB-MP. {\em Top Panel: } The black
dots represent the points in the 20 pixels grid that are closer to
90\% of the peak value; the solid ellipse is the best fit. {\em
Bottom Panel} Same as above, but this time with the 50 pixel grid:
the final ellipse fit is slightly different.}

\label{nodi}  
\end{figure}

To describe in a more quantitative way the shape and structure of the
sub-populations defined in Section~\ref{sample}, we fitted ellipses to
the isodensity contours of each sample. The spatial distributions were
derived as in Section~\ref{dens-par}, with the same kernel, but using
higher resolution grids of 20 and 50 pixel cells. We used the finer
grid (20~pix) for the internal regions that are more densely populated,
i.e., within the 0.6 isodensity level, and the coarser grid (50~pix)
for the outer regions. As in Section~\ref{dens-par}, we defined
densities in units of maximum (peak) density for each distribution, and
we chose isodensity levels going from 90\% to 10\% of the peak density,
in steps of 10\%. Ellipses were fitted to each isodensity line, defined
by those grid nodes that bracket the chosen density value (see
Figure~\ref{nodi}). The best-fit ellipses are hereafter designated with
the same notation used for the isodensity contours in
Figure~\ref{dens}, i.e., 0.9 for the 90\% ellipse and so on.


\citet{kolopov} suggested that the best way to approximate equal 
density lines with ellipses in globular clusters is to use polar 
coordinates ($r$,$\psi$). This method has the advantage that the 
deviations of points from the ellipse along the radial direction are
close to the deviations in the direction perpendicular to the ellipse.
In polar coordinates, the ellipses have the following form:

\begin{equation} 
\frac{1}{r^2}=A+B\cdot \sin{2\psi}+C\cdot\cos{2\psi} 
\end{equation} 
 
\begin{figure} 
\vspace{8cm} 
\includegraphics{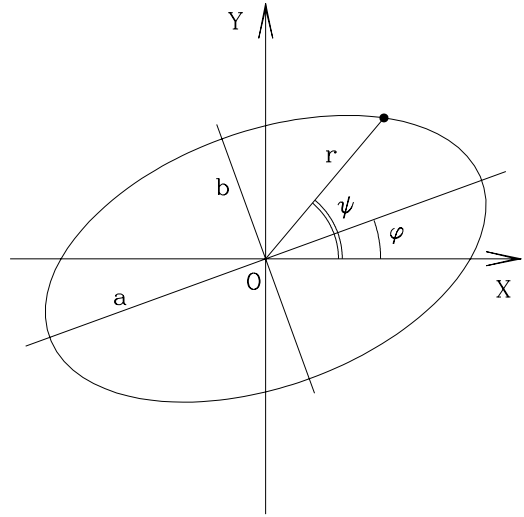} 

\caption{Ellipse parameters notation: $O$ is the ellipse centre; $X$
and $Y$ are the axes of the usual cartesian reference frame, while
$\psi$ and $r$ are the angular and radial coordinates in the polar
coordinate system; $a$ and $b$ are the ellipse semi-major and
semi-minor axes, respectively; $\varphi$ is the major axis inclination
with respect to the $X$ axis.} 

\label{ellisse} 
\end{figure}

The relations of the coefficients $A$, $B$ and $C$ with the usual 
ellipse parameters (Figure~\ref{ellisse}) are
 
\begin{displaymath} 
A=\frac{1}{2}\left(\frac{1}{a^2}+\frac{1}{b^2}\right)\\ 
\end{displaymath} 
\begin{displaymath} 
B=\frac{1}{2}\left(\frac{1}{a^2}-\frac{1}{b^2}\right)\cdot 
\sin{2\varphi}\\ 
\end{displaymath} 
\begin{displaymath} 
C=\frac{1}{2}\left(\frac{1}{a^2}-\frac{1}{b^2}\right)\cdot \cos{2\varphi} 
\end{displaymath} 
 
As demonstrated by earlier studies \citep*{g83}, the ``a priori''
adoption of the cluster centre position can produce errors (i.e., an
overestimation) on the ellipticity estimate. This point is even more
important in our particular case, since we suspect that the centroids
of the three sub-populations differ from each other. Therefore, we
determined the coordinates of each ellipse centre as the mean
coordinates of the input points on each isodensity line. The ellipse
coefficients $A$, $B$ and $C$ where determined by least square
approximation with singular value decomposition \citep{numerical}. An
example of the result of the fitting procedure is shown in
Figure~\ref{nodi}.


\subsection{Ellipse centres} 

Following the procedure described above, we determined the ellipse
centres for each of the three sub-populations defined in
Section~\ref{sample}, and for each isodensity level. The same procedure
has been applied, for ease of comparison, also to the total RGB sample,
defined as the union of the three sub-samples.  The results are listed
in Table~\ref{centres}: the RGB-MInt and RGB-a centres are
significantly different from the RGB-MP centre. We can also compare
with the centre position found by \citet{p00}, which is
(700$\pm$20,1900$\pm$20) in the WFI pixel system. As we can see, the
centre of the global RGB sample is perfectly compatible with that
estimate.

\begin{figure} 
\vspace{8cm} 
\includegraphics{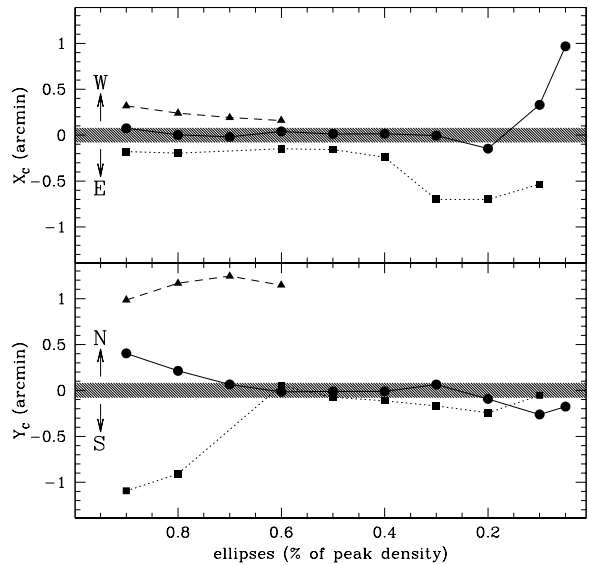} 

\caption{Displacement of the ellipse centres for the
three-sub-populations. Ellipses are denominated along the abscissae in
fractions of the peak value, as described in the text. The shaded area
represents the cluster centre position determined by Pancino et al.
(2000), ($X_C=700\pm 20$~pix, $Y_C=1900\pm 20$~pix). The three
sub-populations are marked with different symbols: {\it filled dots}
for the RGB-MP, {\it filled squares} for the RGB-MInt and {\it filled
triangles} for the RGB-a. The {\it top panel} shows the displacements
in the E-W direction, i.e., along the X CCD axis, while the {\it bottom
panel} shows the displacements along the N-S direction, i.e., the Y CCD
axis. The errorbars are smaller than the symbols.} 

\label{centri} 
\end{figure}

A look at the centre position trend with the isodensity level
(Figure~\ref{centri}) shows that the RGB-MP, dominating the cluster
population, has a quite stable centre position and coincides reasonably
well with the centre of \ome\ measured by \citet{p00}, except for its
two inner isodensity levels. The RGB-MInt centre is slightly displaced
to the east ($\sim10\arcsec$) and significantly to the south
($\sim1\arcmin$), at least for the inner fits. This reflects the
behaviour seen in Figure~\ref{dens}: in the central parts the main peak
of the population is clearly S-E of the RGB-MP population, while the
behaviour in the external parts becomes smoother. The RGB-a centre is
instead slightly displaced to the west ($\sim10\arcsec$), and
significantly to the north ($\sim1\arcmin$), at least for the few
isodensity contour lines that we were able to fit with ellipses. Again,
this reflects what seen in Figure~\ref{dens}. We recall here that the
literature estimates of \ome's core radius go from a maximum of
$r_c=$2.89$\arcmin$ \citep{trager} to a minimum of $r_c=$1.4$\arcmin$
\citep{vanleeuwen}, so the observed displacements are comparable in
size to the core radius of \ome.


\begin{table}

\caption{The density peaks of the whole RGB sample and of the three
sub-samples (in pixels), represented by the centres of the 80\%
isodensity level fits. The distance of each population from the RGB
total sample is calculated, both in pixels and in arcseconds. The last
row reports the corresponding distance between the present RGB global
sample and the cluster centroid determined by \citet{p00}.}

\label{centres} 
\begin{tabular}{lllll} 
{\bf Population} & $X_C$(pix) & $Y_C$(pix) & $d$(pix) & $d$($^{\prime\prime}$) \\ 
\hline
 RGB-tot  & 689$\pm$8  & 1906$\pm$12 &     ---    &    ---   \\ 
 RGB-MP   & 701$\pm$9  & 1954$\pm$17 &  49$\pm$24 & 12$\pm$6 \\ 
 RGB-MInt & 651$\pm$10 & 1671$\pm$16 & 238$\pm$24 & 57$\pm$6 \\ 
 RGB-a    & 761$\pm$9  & 2194$\pm$8  & 297$\pm$19 & 71$\pm$5 \\ 
 \ome     & 700$\pm$20 & 1900$\pm$20 &  12$\pm$32 &  3$\pm$8 \\
\hline 
\end{tabular} 
\end{table} 


\subsection{Flattening and orientation} 
  
As shown in Table~\ref{elli-tab}, the three sub-populations axial
ratios have similar behaviours, within the errorbars. They do not show
dramatic trends moving away from the centre, with both the RGB-MP and
the RGB-MInt becoming slowly rounder away from their density peaks. For
the RGB-a, due to the low number of objects in the external parts,
only  the inner isodensity levels could be fitted. The weighted
averages of the axial ratios shown in Table~\ref{elli-tab} are:
$<(b/a)>=0.81\pm0.01$ for the RGB-MP, $<(b/a)>=0.81\pm0.06$ for the
RGB-MInt and  $<(b/a)>=0.78\pm0.11$ for the RGB-a.

\begin{figure} 
\vspace{4.5cm} 
\includegraphics{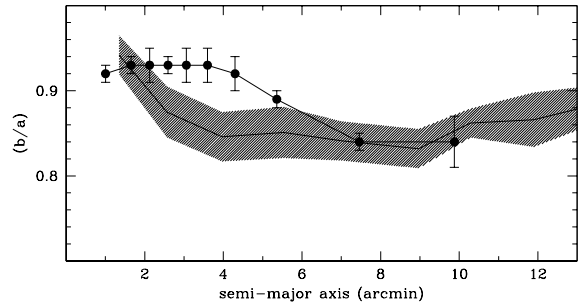}
 
\caption{Axial ratio for the whole RGB population (black dots),
compared with the measures by Geyer et al. (1983), which is represented
by a shaded area (see both their Table~4 and our
Table~\ref{elli-tab}).}

\label{elli-fig} 
\end{figure}

To compare our results with previous work, we performed the same
measurements on the entire RGB sample, resulting from the union of the
three sub-samples defined in Section~\ref{sample}. In particular,
Figure~\ref{elli-fig} shows the comparison with Geyer et al. (1983 --
see their Table~4): the overall agreement is reasonably good, with a
marginal discrepancy in the region between 2$\arcmin$.5 and
4$\arcmin$.5 from the cluster centre. Moreover, our average axial ratio
and ellipticity ($\varepsilon=1-b/a$) for the whole RGB sample,
$<(b/a)>=0.89\pm0.04$ and $<\varepsilon>=0.11\pm0.04$, compare well
with previous results like $<(b/a)>=0.83\pm0.03$ \citep{w&s},
$<\varepsilon>=0.12\pm0.04$ \citep{g83}, or $<\varepsilon>=0.077$
\citep{vanleeuwen}. 

Finally, while the flattenings of the three sub-populations appear
rather similar, the orientations of the best-fit ellipses are instead
quite different (see Table~\ref{elli-tab}). The RGB-MP major axis is
always close to the E-W direction, with $<\varphi
>\sim-4^{\circ}\pm10$, and its inclination seems to increase slowly
with the distance from the cluster centre, although at these low
flattenings the uncertainty on the inclination angle can be
substantially higher than the formal errors quoted in
Table~\ref{elli-tab} \citep{g83}. The RGB-MInt shows instead a complex
structure in the central parts (Figure~\ref{dens}), with a possible
double peak. But its 0.6--0.4 isodensity levels (the most reliable
ones) are always inclined along the N-S direction, with $<\varphi
>\sim100^{\circ}$, while the outer parts suddenly drop to $<\varphi
>\sim25^{\circ}$, much closer to the E-W direction. Finally, the RGB-a
is oriented along the N-S direction, although ellipses could be fit
only to the inner parts, with $<\varphi >$ close to
90$^{\circ}$--100$^{\circ}$, except for the innermost fit. Again, these
quantitative estimates closely reflect what seen in Figure~\ref{dens}.


\section{Metallicity segregation} 
\label{segr-par} 

Recently \citet{j98}, using a compilation of good-quality spectroscopic
data from the literature, has shown that the bright giants
($V\leq12.75$) in \ome, belonging to the two metallicity groups with
[Fe/H]$\leq-1.75$ and [Fe/H]$\geq-1.25$, have a weird spatial
distribution. Each of the two groups occupies one half of the cluster:
no star of one group is found on the other side and the two centroids
are separated by $6.2'$ (see her Figure~1). The separation line runs
approximately perpendicular to the apparent minor axis of the cluster,
which is roughly oriented towards the Galactic centre: the most metal
rich giants are on the southern half, that faces the Galaxy. 

\citet{ikuta} questioned this effect. They used the same compilation
of data and showed that no spatial segregation of stars with different
metallicity is evident. However, we would like to note here that, as
clearly stated by \citet{j98}, the effect is only visible if one
applies the specified selections, i.e., {\it only for bright stars of
the two extreme metallicity groups}. The effect was instead partially
confirmed by \citet{hk00}, who measured abundances of stars in \ome\
with Str\"omgren metallicity indexes. Their Figure~15 shows a clear
segregation of the most metal-rich stars in the southern half of the
cluster, but they could not confirm the segregation of the most
metal-poor stars.


\begin{table} 

\caption{Axial ratio and orientation for the best-fit ellipses for the
whole RGB sample and for each sub-sample. Each best-fit ellipse is
designated with the fraction of the peak density for each population,
as described in the text. The axial ratio $(b/a)$ and the inclination
angle $\varphi$ (i.e., the angle in degrees, counted from west to
north) are shown.}

\label{elli-tab} 
\begin{tabular}{lclcr} 
{\bf Population} & grid & ellipse & $(b/a)\pm\delta(b/a)$ & $\varphi\pm\delta\varphi$ \\ 
\hline 
 RGB-tot  & 20 & 0.9  & 0.92 $\pm$ 0.01 &  49.2 $\pm$  5.7 \\
          & 20 & 0.8  & 0.93 $\pm$ 0.01 &  26.3 $\pm$  4.9 \\
	  & 20 & 0.7  & 0.93 $\pm$ 0.02 &  17.5 $\pm$  3.5 \\ 
	  & 20 & 0.6  & 0.93 $\pm$ 0.01 &  19.5 $\pm$  2.9 \\ 
	  & 50 & 0.5  & 0.93 $\pm$ 0.02 &  18.0 $\pm$  4.3 \\ 
	  & 50 & 0.4  & 0.93 $\pm$ 0.02 &  20.9 $\pm$  4.7 \\ 
	  & 50 & 0.3  & 0.92 $\pm$ 0.02 &  31.9 $\pm$  3.3 \\ 
	  & 50 & 0.2  & 0.89 $\pm$ 0.01 &  32.0 $\pm$  2.7 \\ 
	  & 50 & 0.1  & 0.84 $\pm$ 0.01 &   3.7 $\pm$  1.7 \\ 
	  & 50 & 0.05 & 0.84 $\pm$ 0.03 &   6.0 $\pm$  1.2 \\ 
\hline 
 RGB-MP   & 20 & 0.9  & 0.82 $\pm$ 0.05 & -10.4 $\pm$ 3.1 \\ 
          & 20 & 0.8  & 0.80 $\pm$ 0.12 &   3.8 $\pm$ 2.4 \\
          & 20 & 0.7  & 0.79 $\pm$ 0.05 &   7.0 $\pm$ 1.7 \\
          & 20 & 0.6  & 0.81 $\pm$ 0.04 &   7.0 $\pm$ 1.4 \\
          & 50 & 0.5  & 0.84 $\pm$ 0.03 &  12.4 $\pm$ 2.6 \\
          & 50 & 0.4  & 0.85 $\pm$ 0.02 &  15.8 $\pm$ 2.7 \\
          & 50 & 0.3  & 0.88 $\pm$ 0.02 &  22.7 $\pm$ 3.2 \\
          & 50 & 0.2  & 0.90 $\pm$ 0.01 &  41.9 $\pm$ 2.8 \\
          & 50 & 0.1  & 0.85 $\pm$ 0.04 &   8.6 $\pm$ 2.2 \\
          & 50 & 0.05 & 0.85 $\pm$ 0.02 &  10.9 $\pm$ 1.6 \\
\hline	  
 RGB-MInt & 20 & 0.9  & 0.85 $\pm$ 0.16 &  -8.0 $\pm$  8.9 \\ 
          & 20 & 0.8  & 0.73 $\pm$ 0.10 &  -6.1 $\pm$  2.4 \\
          & 20 & 0.6  & 0.75 $\pm$ 0.02 & 100.2 $\pm$  0.8 \\
          & 50 & 0.5  & 0.77 $\pm$ 0.03 & 100.8 $\pm$  1.6 \\
          & 50 & 0.4  & 0.87 $\pm$ 0.03 & 102.2 $\pm$  2.8 \\
          & 50 & 0.3  & 0.90 $\pm$ 0.02 &  33.0 $\pm$  6.4 \\
          & 50 & 0.2  & 0.88 $\pm$ 0.02 &  24.3 $\pm$  3.8 \\
          & 50 & 0.1  & 0.83 $\pm$ 0.06 &   6.3 $\pm$  2.2 \\
\hline	  
 RGB-a    & 20 & 0.9  & 0.66 $\pm$ 0.11 &  47.2 $\pm$ 10.7 \\ 
          & 20 & 0.8  & 0.86 $\pm$ 0.12 &  97.9 $\pm$  7.2 \\ 
          & 20 & 0.7  & 0.81 $\pm$ 0.04 & 112.7 $\pm$  4.7 \\
          & 20 & 0.6  & 0.83 $\pm$ 0.03 & 108.7 $\pm$  2.9 \\
\hline 
\end{tabular} 
\end{table} 


The metal-rich group in \citet{j98}, with [Fe/H]$\geq-$1.25, is mainly a
sub-sample of our RGB-MInt population, which has a pronounced peak just
$\sim 1^{\prime}$ south of the cluster centre (see Figure~\ref{dens}
and Table~\ref{centres}), a value that is roughly compatible with
Jurcsik estimate ($\sim 3^{\prime}$). We thus easily explain the
observed segregation of her metal-rich group, since when one draws
randomly a sample of metal-rich stars, there is a higher probability to
find them close to the main peak of the density distribution, i.e., in
the southern half of the cluster. 

\citet{j98} metal-poor group, with [Fe/H]$\leq-$1.75, is instead a
sub-sample of our RGB-MP, or more precisely it represents the lowest
metallicity tail of the RGB-MP. We tried to isolate this
sub-population by dividing our RGB-MP sample vertically, in the
($I$,$B-I$) plane, in three sub-samples, equally wide in colour: {\it
(i)} the bluest RGB-MP sub-sample, the RGB-MP1, most probably
corresponding to Jurcsik's metal-poor group; {\it (ii)} the
intermediate sub-sample, the RGB-MP2 and {\it (iii)} the reddest
sub-sample, the RGB-MP3. A two-dimensional KS test, like the one
described in section~\ref{dens-par}, gives a low probability that
the RGB-MP1 is extracted from the same parent distribution of the
RGB-MP2 ($P_{12}=$1.42~10$^{-6}$) or RGB-MP3
($P_{13}=$3.92~10$^{-5}$). However, we were not able to measure any
significant difference in the peaks positions or in the concentration
of the RGB-MP sub-samples. 

We thus are unable to explain the metal-poor part of the spatial
metallicity segregation found by \citet{j98} on the basis of our
photometric catalogue. If \citet{j98} effect will be confirmed and
understood in the future, it could mean that an additional, metal-poor
sub-population exists in \ome, with its own distinct
properties. Otherwise, the observed spatial segregation of that group
of stars could simply be due to a statistical fluctuation, explaining
why \citet{hk00} were not able to confirm the metal-poor part of the
segregation effect. More data on abundances of a significant sample of
stars with [Fe/H]$\leq-1.75$ are thus needed to completely explain
this second half of the puzzle.


\section{Summary and discussion} 
\label{concl}

 Using the nomenclature defined in \citet{p00}, we have selected three
samples with $B<16$: the RGB-MP, the RGB-MInt and the RGB-a. We have
shown that the three samples are significantly  different in their
spatial and structural properties. In particular: 

\begin{enumerate} 

\item{Both the RGB-MInt and the RGB-a main density peaks are shifted 
by $\sim$1$^{\prime}$ with respect to the cluster centre.} 

\item{Both the RGB-a and the RGB-MInt populations are elongated on a 
direction that is perpendicular to the elongation of the RGB-MP 
population.}

\item{The RGB-a population is significantly more concentrated than the
RGB-MP population. No firm conclusion can be reached for the RGB-MInt,
due to the complex isodensity contour line's shapes in its central
parts.} 

\item{Both the RGB-MInt and the RGB-a show complex and perturbed 
isodensity contour lines, resembling bubbles, shells and tails. While
for the outer isodensity contours this can be due to statistical
undersampling (especially for the RGB-a), in the inner parts these
features are probably real.}

\item{Finally, we were able to explain the metal-rich part of the weird
spatial metallicity segregation, found by \citet{j98} and confirmed by
\citet{hk00}, thanks to the peculiar surface density distribution of
the RGB-MInt and, in particular, to its centre displacement with
respect to the RGB-MP centre.} 

\end{enumerate} 

Let us discuss these findings in the framework of the current scenarios
for the formation and evolution of \ome. As briefly mentioned in
Section~1, there is a consistent body of evidence supporting the fact
that \ome\  built up by itself at least part of the chemical elements
that we can observe today. In a standard self-enrichment scenario, we
would expect the more metal-rich (and younger) populations to be more
centrally concentrated than the metal-poor, dominant population. This
is almost always the case for the dwarf galaxies of the local group,
with only a few exceptions \citep{grebel}, and it is exactly what we
observe for \ome\  (point (iii) above).

Unfortunately, while there is no doubt on the fact that
self-enrichment must be one of the fundamental ingredients of a
successful formation and evolution theory for \ome\
\citep{norris,smith}, we are dealing with a complex set of
observational facts, containing conflicting evidence. In fact, a few
of the observational properties of \ome, concerning the structure, the
shape and the kinematics (Section~\ref{intro}) are not easily
accommodated into a simple self-enrichment scenario
\citep{n97,p00,f02}, and suggest the possibility of a merger or
accretion event. However, the simple merging of two or more single
metallicity clusters cannot account for the broad metallicity
distribution of the RGB \citep{n96,smith} and the high speed of
ordinary, already formed globular clusters in the present potential
well of the Milky Way makes this kind of merging quite unlikely
\citep{thurl}\footnote{In a lower potential environment, like in the
Fornax dwarf spheroidal, the chance would be substantially higher.}.
The evidence presented here, concerning the structure and shape of the
different sub-populations in the RGB of \ome, supports these findings
and thus confirms the need for a more sophisticated scenario, that
takes into account {\em all} the observational evidence collected so
far.

\subsection{A Complex Dynamical History}

Let us first discuss the relative orientation of the three RGB
sub-populations. It is now well established that the elongated shape
of the whole cluster (dominated by the RGB-MP population) is mainly
due to rotation \citep{harding,merritt94,meylan}, which is consistent
with the picture of a dynamically young and not completely relaxed
cluster: the cluster's peak rotational velocity is 7~km~s$^{-1}$ at
11~pc from the centre \citep{merritt97} and the relaxation times for
\ome\ are of the order of magnituide of its age, being $\log
t_{rc}=$9.73~yr in the core and $\log t_{rh}=$10~yr at half mass
\citep{harris}.

Thus, it becomes tempting to explain the elongations of the RGB-MInt
and RGB-a populations in terms of rotational velocity, too, but in this
case it would be necessary to assume that these two populations rotate
around a {\it perpendicular axis} with respect to the RGB-MP. An
inspiring comparison is posed by the recent work by Sarzi et al.
(2000), who examine an example of galaxy that has undergone a major
merging or accretion event in its past. The signature of such an event
is the simultaneous presence of {\it (1)} an orthogonally elongated
bulge with respect to the disk, and of {\it (2)} two rotation curves,
perpendicular to each other, one for the host galaxy and another for 
the accreted component. 

Now, we have shown here (point (ii) above) that the first signature
could indeed be present in \ome: we have (at least) two components
with orthogonal elongations. What can we say about the rotation
curves? It has been demonstrated in the past \citep{n97} that while
the metal-poor stars in \ome\ rotate, no sign of rotation is evident
for the metal-rich stars. Again, the metal-rich stars of Norris
correspond mainly to our RGB-MInt population, while too few data are
presently available for the RGB-a population. So, the second signature
is only partially present in \ome. Clearly, to fully demonstrate or
falsify the point, one needs more precise radial velocities for a much
larger sample of RGB-MInt and RGB-a stars, and a model which is best
suited to small systems embedded in a common potential well. For
example, if we consider simulations of the merger of two globular
clusters with unequal masses \citep{makino}, the most massive object
retains a larger share of the initial orbital angular momentum, as
\citet{n97} point out to explain why we see rotation for the
metal-poor stars and not for the metal-rich ones.

\subsection{An Accreted Component?}

Another interesting point has recently emerged from the coupling of
the photometric catalog by \citet{p00} and the proper motion work by
\citet{vanleeuwen}, i.e., that the RGB-a population appears to have a
different bulk proper motion of $\mid\delta\mu\mid$=0.8~mas~yr$^{-1}$
with respect to the main population of \ome\ \citep{f02}. This
evidence suggests that the RGB-a could be an accreted population,
captured by the main body of \ome. We thus should expect the RGB-a to
have a different centre from the main cluster population, too, and it
is exactly what we find here (point (i) above). Moreover, if a merging
event has really taken place in the past history of \ome, how long ago
do we expect it to have happened? We already know that \ome\ has a
very long relaxation time, comparable to its age. We also know that
the RGB-a is $\sim$2--3 times more concentrated than the RGB-MP (point
(iii) above), and the fact that it is almost one core radius away from
the centre suggests that it is moving in a significantly less dense
environment, so it could have survived as a self-gravitating system
for many Gyr. It is however puzzling that the (few) radial velocities
measured for the RGB-a so far are not so different from the cluster
average\footnote{In this respect, it should be noted that we could be
seeing the RGB-a ``orbit'' from its pole, although this configuration
does not have a large probability of being observed.}.

So, a self-consistent set of observational facts, supporting the
occurrence of an accretion or merger event in the cluster past history,
is beginning to take shape. We would like to note here that this fact
in not necessarily in contradiction with the occurring of
self-enrichment, nor with the hypothesis that \ome\  is the remaining
nucleus of a larger body (a dwarf galaxy), accreted and partially
disrupted by the Milky Way. If we consider the case of the Sagittarius
dwarf spheroidal (Sgr), which is the only one showing some resemblance
with the case of \ome, we see that more than one GC seems to be
associated with the system. In particular, one of these (Terzan~7) has
a metallicity of [Fe/H]$\sim$--0.62 \citep{harris} compared with the
[Fe/H]$\sim$--1.59 of M~54, which may be the nucleus of the Sgr.
Interestingly, the RGB-MP has a similar metallicity to that of M~54,
and the RGB-a to Terzan~7\footnote{An interesting discussion on the
survival of globular clusters within dwarf elliptical galaxies has been
presented by \citet{lotz}.}. 

\subsection{A Promising Framework}

An interesting scenario, that could accommodate all of the
observational evidence collected so far, was discussed by, e.g.,
\citet{freeman}, \citet{n97} and \citet{smith}: the so-called {\it
merger within a fragment scenario}, descending from the general
framework proposed by \citet{s77} and \citet{searle}. In this
framework, a conglomerate of star-forming sub-systems or regions could
evolve within a large cloud and a common potential well (a fragment),
each section evolving with slightly different timescales, and slightly
different chemical properties. In this context, the chemical evolution
of the RGB-MP and of the RGB-MInt could have been tightly related to
each other, especially since there are reasons to believe that the
RGB-MInt is younger by a few Gyr \citep{hughes} and could have been
enriched by the ejecta of the RGB-MP stars.

Later, a dwarf galaxy with its own globular cluster system could form,
and a few globular clusters (like the RGB-a, or even the RGB-MInt)
could spiral towards the system centre, while the remaining clusters
could be stripped by the interaction with the Milky Way, along with
most of the dwarf galaxy halo. Recent calculations by \citet{bromm}
support this line of reasoning \citep[see also][]{kroupa}, together
with the example of the Sagittarius dwarf galaxy discussed above.

Although still speculative, this idea deserves to be further explored
since it appears the most promising to explain \ome's puzzling
properties.


\section*{Acknowledgments}

We would like to thank E. Pignatelli and E. Pompei for interesting
discussions on galaxy mergers. The financial support of the {\it
Agenzia Spaziale Italiana} (ASI) is kindly acknowleged. This research
was also partially supported by the Italian {\it Ministero
dell'Istruzione dell'Universit\`a e della Ricerca} (MIUR). E. Pancino
is grateful to the Eropean Southern Observatory (ESO), where part of
this work was carried out, within the {\it Studentship Programme}. A.
Seleznev acknowledges the support of the Russian Federation State
Scientific and Technical Program {\it ``Astronomy''}.


\label{lastpage} 
 
\end{document}